\documentclass[%
 aip,
 amsmath,amssymb,
 reprint,%
]{revtex4-1}

\usepackage{graphicx}
\usepackage{dcolumn}
\usepackage{bm}

\usepackage[utf8]{inputenc}
\usepackage[T1]{fontenc}
\usepackage{mathptmx}
\usepackage{etoolbox}

\makeatletter
\def\@email#1#2{%
 \endgroup
 \patchcmd{\titleblock@produce}
  {\frontmatter@RRAPformat}
  {\frontmatter@RRAPformat{\produce@RRAP{*#1\href{mailto:#2}{#2}}}\frontmatter@RRAPformat}
  {}{}
}%
\makeatother
\begin{document}

\preprint{AIP/123-QED}

\title[Polarization‑Sensitive Third Harmonic Generation in resonant silicon nitride Metasurfaces for deep-UV Emission]{Polarization‑Sensitive Third Harmonic Generation in resonant silicon nitride Metasurfaces for deep-UV Emission}
\author{S. Mukhopadhyay}
\altaffiliation{Corresponding author}
\affiliation{Department of Physics – Universitat Politècnica de Catalunya, Rambla Sant Nebridi 22, 08222 Terrassa (Barcelona), Spain.}

\email{shroddha.mukhopadhyay@upc.edu}

\author{M. A. Vincenti}
\affiliation{Department of Information Engineering – Università degli Studi di Brescia, Via Branze 38, 25123 Brescia, Italy}%

\author{R. Malureanu}
\affiliation{DTU Nanolab, National Center for Nanofabrication and Characterization, Oersteds Plads, building. 347, Dk-2800, Kgs. Lyngby}
\affiliation{DTU Electro, Department of Electrical and Photonics Engineering, Oersteds Plads, building.343, Dk-2800, Kgs. Lyngby}

\author{C. Cojocaru}
\affiliation{Department of Physics – Universitat Politècnica de Catalunya, Rambla Sant Nebridi 22, 08222 Terrassa (Barcelona), Spain.}

\author{M. Scalora}
\affiliation{Department of Information Engineering – Università degli Studi di Brescia, Via Branze 38, 25123 Brescia, Italy}

\author{J. Trull}
\affiliation{Department of Physics – Universitat Politècnica de Catalunya, Rambla Sant Nebridi 22, 08222 Terrassa (Barcelona), Spain.}

\date{\today}

\begin{abstract}
We present a combined experimental and theoretical study of enhanced third-harmonic generation (THG) in silicon nitride metasurfaces. These structures exhibit strong resonant nonlinear responses, enabling up to two orders of magnitude enhancement in THG compared to a flat silicon nitride etalon, driven by strong electromagnetic field localization. We investigate two polarization-selective metasurface geometries supporting transverse electric (TE) and transverse magnetic (TM) resonances, implemented in fully planar architecture. When driven by ultrafast near-infrared laser pulses, these resonances confine optical energy at the nanoscale, enabling efficient frequency up-conversion from the visible to the ultraviolet (UV) and deep-UV spectral regions. Through spectral mapping of the nonlinear response under both TE and TM excitation, we quantify field confinement, extract the effective nonlinear enhancement, and characterize the spectral dependence of the third-harmonic generation efficiency. This two-dimensional periodic platform provides a flexible design toolbox for engineering polarization-dependent UV and deep-UV light sources. More broadly, our results demonstrate that silicon nitride, a CMOS-compatible dielectric, can support efficient nonlinear up-conversion deep into the UV. This finding shows that access to short-wavelength nonlinear photonics does not require complex materials or architectures, but can instead be achieved using widely available dielectrics through careful structural design.
\end{abstract}

\maketitle

\begin{quotation}

\end{quotation}

\section{\label{sec1}Introduction}

Recent advances in high index semiconductor materials operating within their transparency windows have enabled a new generation of tunable metasurfaces whose emergence has stimulated extensive research activity and naturally broadened the range of applications in nanophotonics.  Unlike plasmonic systems, which suffer from significant losses due to strong absorption, high‑index dielectric resonators can support similar dynamics with negligible dissipation across their operating bandwidth. They provide strong field confinement and extensive design flexibility, which has enabled a wide range of demonstrations involving chiral light control, coherent wavefront shaping, and topological optical modes \cite{bib1,bib2,bib3,bib4}. Silicon (Si),owing to its natural abundance and mature fabrication ecosystem, has long been the dominant material for integrated electronic and photonic devices  \cite{bib5,bib6,bib7}. Its high refractive index has enabled widespread use in waveguiding platforms \cite{bib8,bib9,bib10,bib11,bib12}, and its large third-order nonlinearity has made it a natural candidate for integrated nonlinear optics \cite{bib13,bib14}. Strong optical confinement in Si waveguides, combined with a high third order nonlinearity, allows efficient frequency conversion in compact geometries. 

On the opposite end of the spectrum, materials like silicon dioxide, typically employed for optical fibers \cite{bib15}, trade transparency for poor nonlinear performance. Silicon nitride ($Si_3N_4$) has recently emerged as a compelling alternative: it is amorphous, CMOS-compatible, moderately high refractive index ($n\approx2$), and exhibits a relatively high intrinsic nonlinear third order susceptibility with very low optical loss. \cite{bib16,bib17}. Its wide bandgap is not susceptible to two-photon absorption at telecom wavelengths, enabling efficient third harmonic generation, optical parametric oscillation, self-phase modulation, and broadband supercontinuum generation  \cite{bib18,bib19,bib20,bib21}. These capabilities underpin a broad range of applications, including data communications and telecommunications \cite{bib22,bib23}, biosensing \cite{bib24}, positioning and navigation \cite{bib25}, low noise microwave synthesizers \cite{bib26}, spectroscopy \cite{bib27}, radio-frequency (RF) signal processing \cite{bib28}, quantum communication \cite{bib29}, and atomic clocks \cite{bib30}. To date, however, most $Si_3N_4$ based nonlinear studies have focused on integrated waveguide components \cite{bib18,bib19,bib20,bib21} supported by rapidly advancing fabrication processes \cite{bib31,bib32,bib33} and primarily operating in the near-infrared and telecom wavelength range \cite{bib20,bib34,bib35,bib36,bib37}. Despite significant progress in fabricating 2D and 3D nanostructures and metamaterials, theoretical efforts have largely concentrated on linear optical behavior.  As a result, substantial work remains to fully understand and exploit nonlinear interactions in these structured platforms, particularly under high-intensity, or ultrashort excitation.

Here we investigate the third order nonlinear optical properties of $Si_3N_4$, probed experimentally using high-power femtosecond laser pulses and quantified by the generated third harmonic signal from two distinct nanogratings, etched into the same $Si_3N_4$ membrane. Although a simple, subwavelength thick membrane offers only limited interaction lengths and intrinsically low conversion efficiencies, the investigation of its nonlinear properties is a necessary step that provides a clean platform for extracting the intrinsic nonlinear parameters of the material. Establishing these parameters first is essential: it reveals the fundamental capabilities and limitations of the material, in this case $Si_3N_4$, such as the presence of absorption resonances or native field‑enhancement channels and determines whether the material’s baseline nonlinear response is strong enough to justify pursuing more complex nanostructured geometries for artificial enhancement. We report the theoretical and experimental observation of third harmonic generation from two kinds of polarization sensitive $Si_3N_4$ metasurface designs: a fully suspended nanograting and a partially etched grating into the membrane. These gratings  strongly confine the electromagnetic field under resonant illumination, thereby enhancing light matter interaction at the nanoscale, resulting into an impressive two orders of magnitude enhancement of the THG signal, relative to the flat $Si_3N_4$ membrane. Our study focuses on visible wavelength excitation to generate significantly enhanced THG in the deep-UV, illustrating how these structures can serve as efficient UV and deep-UV light sources. By combining experiments, theory, and full-wave simulations, we obtain a detailed understanding of the resonant dielectric response and the underlying nonlinear electron dynamics. Key advantages of these platforms include: polarization-selective operation; enhancing both TE and TM inputs without additional optics; and tunable wavelength conversion from the visible to deep-UV regimes by adjusting the angle of incidence.   

\section{Silicon Nitride resonant structures : designs and linear properties}
\label{sec2}

\textit{Free standing $Si_3N_4$ membrane}

\begin{figure*}[hbt!]
    \centering
    \includegraphics[width=.8\linewidth]{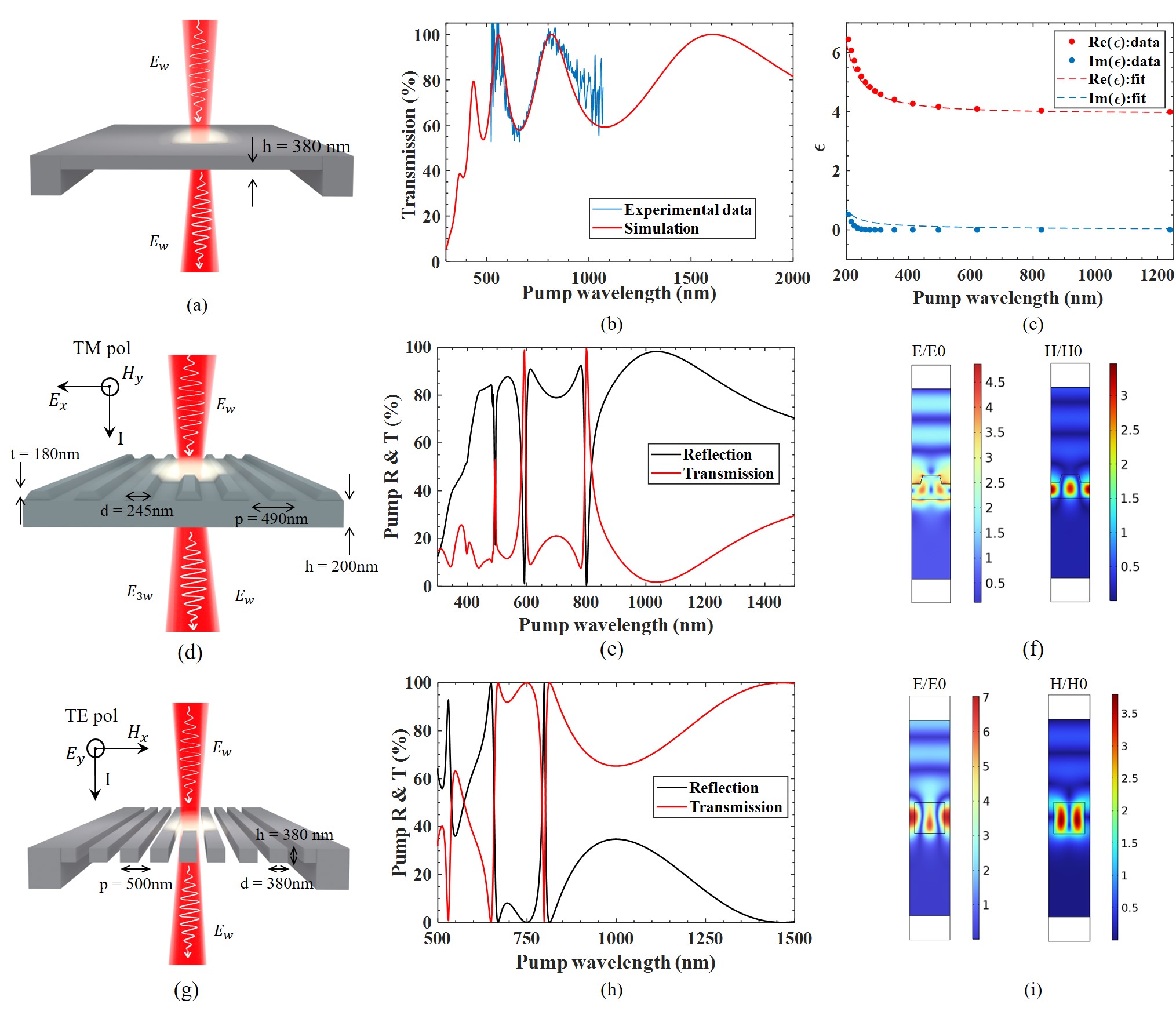}
    \caption{(a) top to bottom: Schematic representation of $Si_3N_4$ membranes under study, (b) Linear transmission through a 380 nm-thick membrane, (c) Dielectric permittivity function of $Si_3N_4$ extracted from Palik’s handbook, and Lorentzian fit. (d) Schematic representation of the partially etched grating under study (e) Linear transmission through the sample showing our target guided mode resonance $\lambda_{pump}=800nm$ (f) E and H field localization (normalized) at our target resonance at $\lambda_{pump}=800nm$ (g) Schematic representation of the fully etched suspended grating under study (h) Linear transmission through the sample showing our target mie resonance at   (i) E and H field localization (normalized) at our target resonance at  $\lambda_{pump}=800nm$}
    \label{fig1}
\end{figure*}

As a reference, we begin by characterizing a free standing $Si_3N_4$ membrane, shown schematically in Fig.\ref{fig1}(a). The thickness of the membrane has been adjusted to be resonant at 800 nm, a wavelength that we will use as a pump. The theoretical prediction of the transmission versus the wavelength for a 380 nm thick membrane is shown as red curve in Fig.\ref{fig1}(b). For the linear characterization of the transmission, the fabricated membrane is illuminated with a supercontinuum continuous wave (CW) laser source, and the transmitted light is collected with a high-resolution spectrometer. The blue curve in Fig.\ref{fig1}(b) reports wavelength-dependent pump transmittance through the  380 nm-thick membrane, experimentally retrieved and normalized to the incident spectrum. It exhibits a series of Fabry-Peròt resonances whose free spectral range is determined by the membrane thickness. The experimentally observed reflection from the samples were negligible. The agreement between simulations and experiment is excellent, both in amplitude and in the location of the resonance peaks.

Fig.\ref{fig1}(c) shows the linear dielectric response of $Si_3N_4$ as reported in Palik’s handbook (curves with markers), which is fitted using a superposition of three Lorentzian functions to accurately represent the local dielectric response in the wavelength range above 200 nm:

\begin{equation}
\begin{split}
\epsilon (\omega) = & 1 -  \frac{ {\omega}^2_{p1}}{\omega^2 -  {\omega}^2_{01} + i  {\gamma}_{01} \omega} \\
& -\frac{ {\omega}^2_{p2}}{\omega^2 -  {\omega}^2_{02} + i  {\gamma}_{02} \omega} - \frac{ {\omega}^2_{p3}}{\omega^2 -  {\omega}^2_{03} + i  {\gamma}_{03} \omega} 
\end{split}
\label{eqn1}
\end{equation}

where $(\omega{p1},\omega_{01},\gamma_{01})=(7.75,6.5,0.4)$, $(\omega{p2},\omega_{02},\gamma_{02})=(7.85,8.95,2.75)$, $(\omega{p1},\omega_{01},\gamma_{01})=(14.5,16.75,2.95)$. The scaled frequency $\omega=\frac{1}{\lambda}$, where $\lambda$ is expressed in microns; each of the remaining coefficients is scaled in units of $\frac{\lambda_0}{c}$, where $\lambda_0=1\mu m$ . 

\textit{Partially etched grating}

The second configuration we explore is a periodically structured nanoscale grating, etched on top of the aforementioned 380 nm thick freestanding $Si_3N_4$ membrane, that supports guided mode resonances. The partially etched grating was designed (Supplementary info Fig.S1) to attain a characteristic resonance tuned at 800 nm under TM polarized illumination, i.e., electric field of the incident light is perpendicular to the grating orientation [see Fig.\ref{fig1}(d)]. The depth and position of the resonance depend critically on several geometrical parameters: period (p=490 nm), width of the trenches (d=p/2=245 nm), depth of the trenches (t=180 nm), as well as the thickness of the unetched $Si_3N_4$ base (h=200 nm). We achieve the resonance position at 800 nm, for p=490 nm. Fig.\ref{fig1}(e), shows the full spectrum of the simulated transmission curve for the structure, and Fig.\ref{fig1}(f), shows the E and H field confinements at the targeted wavelength of resonance $\lambda_{pump}=800nm$. During design, the fabrication limits were considered in the simulations as the maximum inclination of walls possible to achieve in a partial chemical etching procedure is 75° (Fig.\ref{fig1}(c)).

\textit{Fully etched grating in the membrane}

A complementary route to tailoring electromagnetic mode structure is to exploit a Mie-type resonance. In this regime, the characteristic dimensions of the nano-resonator set the resonant wavelength, and the scattering response can be engineered directly by adjusting the geometry of the meta-atoms. We designed a fully etched, free standing grating suspended in air. This structure supports a pronounced Mie resonance under TE polarized illumination, i.e., when the E field is oriented parallel to the grating bars. [see Fig.\ref{fig1}(g)]Numerical simulations were done to target a resonance tuned to 800 nm for a grating with a period p=500 nm and bridge dimensions h=d=380 nm. Under TE excitation, the E field efficiently couples to a localized displacement current mode inside the high index material, producing a sharp dip in the transmission spectrum (Fig.\ref{fig1}(h)). This spectral feature arises from strong, wavelength selective field confinement within each dielectric bar (Fig.\ref{fig1}(i)). The resulting resonance is significantly narrower than the guided‑mode resonance observed in a partially etched grating, which will be discussed next. This difference reflects the distinct physical origins of the two modes: a Mie resonance is an intrinsic, highly localized, low‑loss eigenmode of each individual scatterer, whereas a guided‑mode resonance is a collective, nonlocal effect enabled by the periodicity of the structure and broadened by radiative leakage and diffraction‑mediated coupling  \cite{bib38, bib39,bib40,bib41}. In both grating platforms, the intense electric‑field localization at resonance substantially enhances the nonlinear light–matter interaction and results in strong third‑harmonic generation at and around the respective resonance wavelengths, as demonstrated in section \ref{sec5}.

\section{Sample fabrication}
\label{sec3}
All three samples were fabricated using aligned electron beam and UV lithography (Fig.2). In a first instance, a 380 nm Si-rich silicon nitride was deposited on both sides of double-side polished, 350 mm thick Si wafers. A Si‑rich silicon nitride composition was chosen to reduce mechanical stress in the films. This approach prevents us from precisely determining the stoichiometric coefficients of $Si_xN_y$. However, the composition is expected to remain close to that of $Si_3N_4$. This assumption is supported by the excellent agreement obtained when fitting the linear dielectric constant with the tabulated $Si_3N_4$ data from Palik.

\begin{figure}[hbt!]
    \centering
    \includegraphics[width=\linewidth]{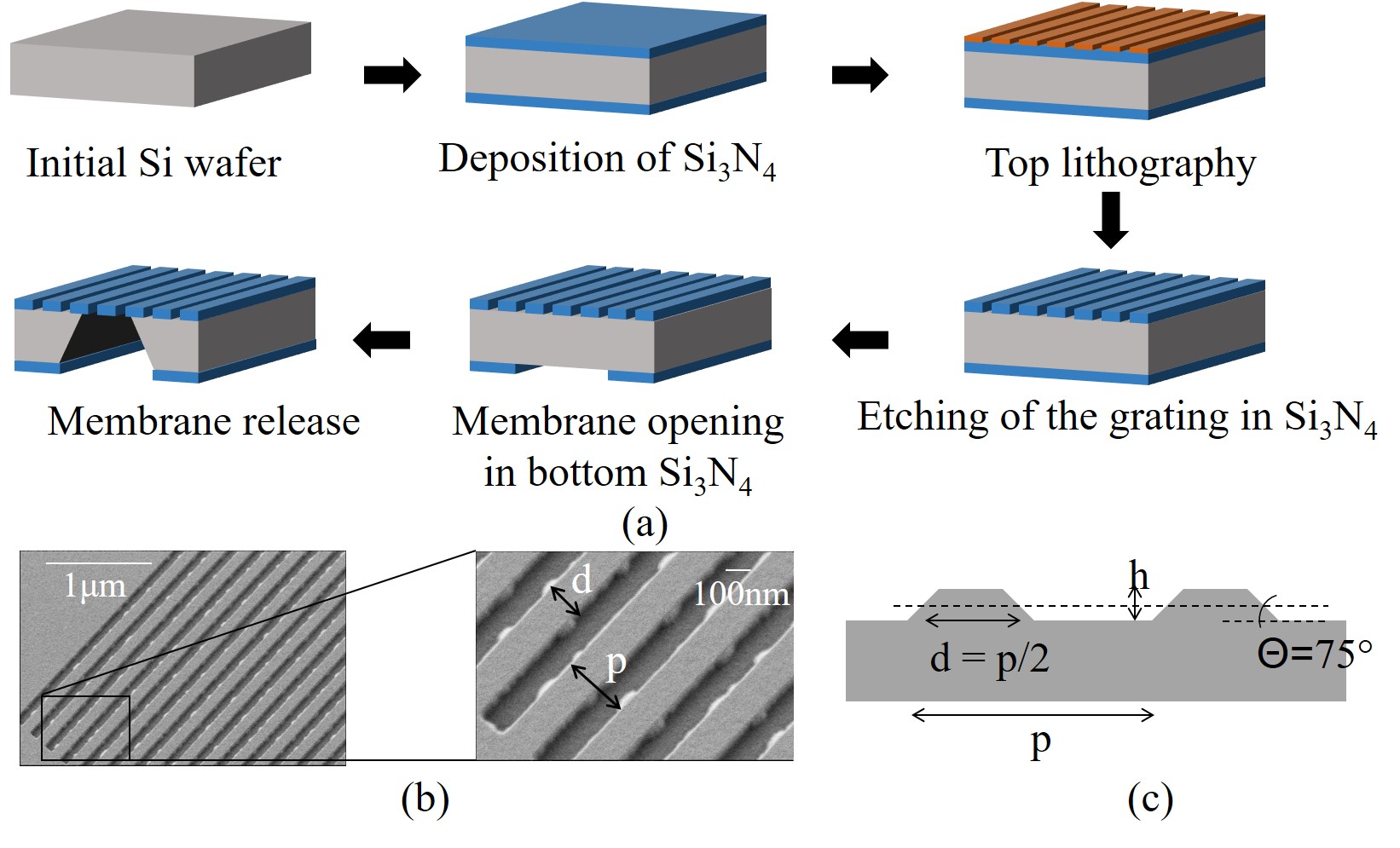}
    \caption{(a) Schematic representation of the fabrication process. (b) Scanning electron microscope (SEM) image of the sample with zoomed view. (c) Side view of the tilted walls of the samples because of partial etching.}
    \label{fig2}
\end{figure}
On one side of the wafer, we performed an electron-beam lithography to define, in resist, the desired geometry shown in Fig.\ref{fig1}(d) and \ref{fig1}(g).The resist was then used as a mask in an ICP plasma etcher to etch the silicon nitride. For the fully etched gratings, in order to obtain as vertical a sidewall as possible, we over etched into the Si beneath. This over etching will not affect the final structure since the Si beneath the grating needs to be removed.

Once the top side with the gratings was finished, we then performed an aligned UV exposure on the other side of the wafer to etch the Si wafer and obtain the silicon nitride membranes.

One of the last processing steps was the Si etching. It was performed in a KOH bath at 80 degrees for 5 hours. The KOH bath etches Si along the [100] crystallographic plane much faster than along the [111] plane thus creating a pyramid-like hole in Si.

Last step was drying. To avoid the collapse of the fully-etched structures (Fig.\ref{fig1}(g)), we took two precautions: first, between each grating line we interposed, at the design and exposure stage, sustaining beams separated by 15 microns. That means that the longest free-standing beam is 15 microns long rather than 200 microns, the whole length of the grating. The second precaution was taken at the end of the process: we dried these samples using a critical point dryer thus avoiding the creation of a gas-liquid interface that might have collapsed the grating onto itself.

\section{Experimental setup for nonlinear characterization}
\label{sec4}

\begin{figure}[hbt!]
    \centering
    \includegraphics[width=\linewidth]{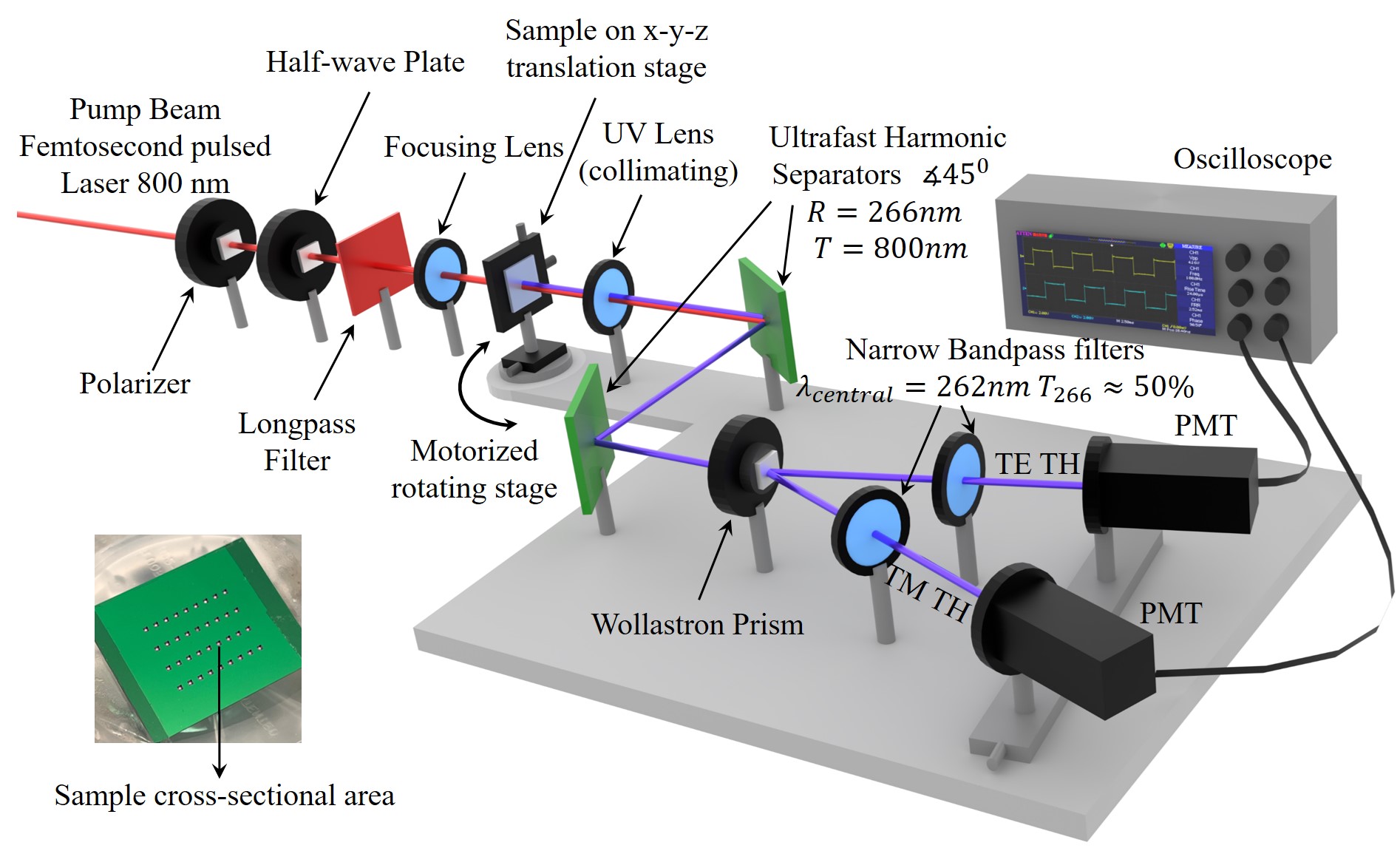}
    \caption{Schematics of the experimental set-up used to measure transmitted third harmonic signals generated by the $Si_3N_4$ samples as a function of angle of incidence, polarization, and incident wavelength. Inset shows one of the chips; each small square contains one sample suspended at its back surface. }
    \label{fig3} 
\end{figure}

To efficiently detect the THG from the nanostructures, we constructed the optical setup illustrated in Fig.\ref{fig3}. The arrangement enables quantitative measurements of the power conversion efficiency from the incident pump to the THG frequency, defined as $\eta_{TH}=\frac{P_{TH}}{P_{PUMP}}$, with detectable values ranging from $10^{-6}$ down to $10^{-10}$ level. 

The pump source for wavelength-dependent nonlinear measurement is a fiber-amplified femtosecond (fs) laser equipped with an optical parametric amplifier (CARBIDE+OPA from Light Conversion), tunable from 650 nm to 2500 nm. The system delivers pulses of approximately 150 fs duration at a 60 KHz repetition rate, with pulse energy up to 6 $\mu$J and a maximum average power of 400 mWatt. The output polarization (TE or TM) is controlled using a calcite Glan-laser polarizer followed by a continuously rotating half-wave plate. A near-IR long-pass filter suppresses any spurious nonlinear signals generated by other optical components along the beam path. The pump beam is focused onto the sample using a plano-convex lens of f=200 mm, yielding peak intensities of the order of $\sim10 GW/cm^2$. Each sample has a square cross-sectional area of $200\mu m×200\mu m$ patterned region, sufficiently large to accommodate the focused beam, which has a diameter of approximately $97\mu m$ (calculated at $\frac{1}{e^2}$ intensity) at normal or near normal incidence. During wavelength sweeps, the pump wavelength was changed manually. For each wavelength, the laser energy is tuned to compensate for the OPA’s wavelength dependent conversion efficiency. In addition, wavelength specific responsivity values were applied to the detection system, and a correction factor was used to account for the wavelength-dependent transmission (or reflection) of all optical elements in the beam path. The samples are mounted on an XYZ translation stage attached to a motorized rotation platform, allowing precise control of the incidence angle with a resolution of 0.2°.  

Fig.\ref{fig3} illustrates the TH detection arm. The TH signal emerging from the sample is collimated using a UV-enhanced, anti-reflection coating lens. Two dichroic mirrors then act as ultrafast harmonic separators: they efficiently reflect the THG wavelength (centered at 266 nm) while transmitting the fundamental pump (centered at 800 nm), operating at 45° angle of incidence. The beam then passes through a Wollaston polarizer, which separates the TM and TE components of the TH field for simultaneous measurement. A narrow bandpass filter centered at 262 nm further isolates the TH radiation by strongly suppressing any residual pump light, followed by a photomultiplier tube (PMT), to detect the resulting signal with high sensitivity. A mechanical chopper placed immediately after the input laser modulates the pump with a square wave profile, enabling lock-in detection and improving signal-to-noise ratio. A detailed calibration procedure, accounting for the transmission of each optical component and the wavelength-dependent responsivity of the PMT, allows us to determine the absolute conversion efficiency as the ratio of the TH energy to the incident pump energy. The entire detection assembly, which is the most critical part of the setup, is mounted on a rotary platform that enables precise control of the detection angle for angular measurements. Extensive alignment and calibration were performed to ensure accurate and reproducible efficiency measurements.

\section{Results and discussion}
\label{sec5}
\begin{figure*}[hbt!]
     \centering
     \includegraphics[width=.9\linewidth]{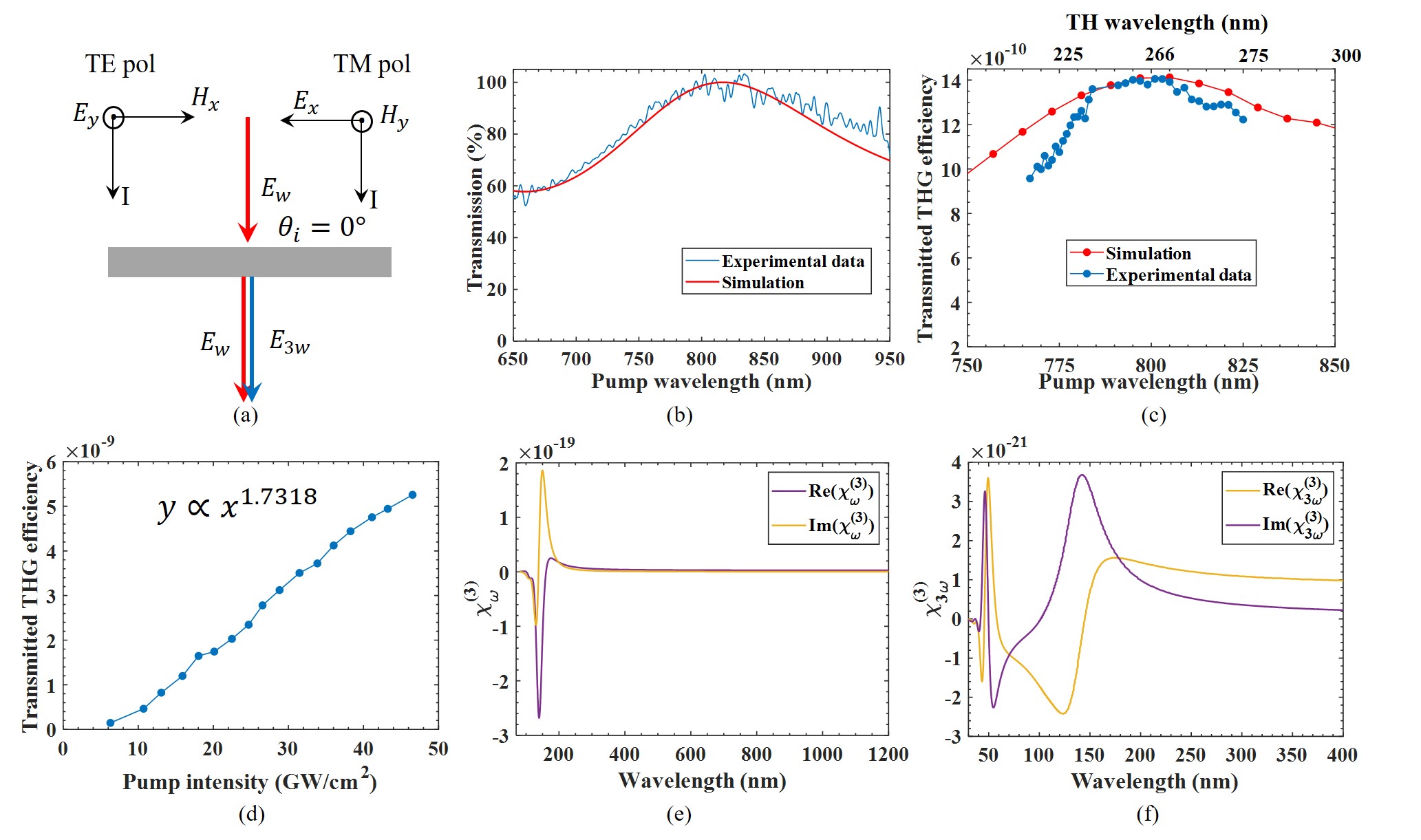}
     \caption{(a) Schematic of the membrane illumination (b) Peak of the Fabry-Peròt resonance (c) Spectral dependence of THG from $Si_3N_4$ membrane (b) Power dependence of THG conversion efficiency for fixed pump wavelength of 800 nm (c)-(d) Wavelength dependent $\chi^{(3)}_\omega$ and $\chi^{(3)}_{3\omega}$ datasets retrieved from the THG measurements.}
     \label{fig4}
 \end{figure*}

In this section we represent the comprehensive results of our THG spectral measurements and analysis from the above mentioned $Si_3N_4$ samples. We begin with the $Si_3N_4$ membrane, its characteristic Fabry-Peròt resonance around 800 nm being reflected into the THG spectral response. In Fig.\ref{fig4}(a) we present a schematic of the membrane and in Fig.\ref{fig4}(b) we represent the peak of the Fabry-Peròt resonance of the membrane, the red curve being the simulated and the blue curve corresponding to the experimentally retrieved spectra. Fig.\ref{fig4}(c) represents the experimentally retrieved THG conversion efficiency curve in blue, with a peak conversion efficiency of $\eta_{TH}=\frac{P_{TH}}{P_{pump}}=1.4\ast 10^{-9}$. The theoretical spectral response of the THG as shown in red curve, is fitted with help of our microscopic, hydrodynamic model for linear and NL optical properties of semiconductors \cite{bib42,bib43,bib44,bib45}. The THG is observed for TM input-TM output (Fig.\ref{fig4}(c)) and TE input-TE output (Supplementary info Fig.S3) polarization states, respectively, and is consistent with the $\chi^{(3)}$ characteristics of isotropic materials, where only one element of the $\chi^{(3)}$  matrix $\chi^{(3)}_{ijkl}(3\omega,\omega,\omega,\omega)$ is independent \cite{bib46}. Fig.\ref{fig4}(d) on the other hand, shows the experimentally retrieved THG conversion efficiency as a function of the incident pump intensity, for a fixed pump wavelength of 800 nm. This was performed to verify that the THG signal intensity is proportional to the cubic power of the pump beam intensity, which, in turn, means that the THG conversion efficiency depends quadratically on the pump beam intensity. Our experimental data shows a y-x dependence of the curve proportional to $\sim x^{1.73}$, which is justified as, under laboratory conditions, a minimal amount of environmental noise is added to the measurements. The above-mentioned measurements allow us to identify, distinguish, and explain the different nonlinear contributions to the harmonics generated by silicon nitride at the nanoscale at visible and UV wavelengths. The aim is to identify the basic physical parameters of $Si_3N_4$, such as the effective mass of bound electrons and nonlinear oscillator parameters, which control the THG, and to make accurate predictions about conversion efficiencies in more complicated geometric arrangements. To fit the THG characteristics of the planar membrane, we retrieve the wavelength dependent datasets of $\chi^{(3)}_\omega$ and $\chi^{(3)}_{3\omega}$ respectively, for our specific pump intensity and fixed laboratory environment [Fig.\ref{fig4}(e) and Fig.\ref{fig4}(f) respectively]. 

\begin{figure*}[hbt!]
    \centering
    \includegraphics[width=.9\linewidth]{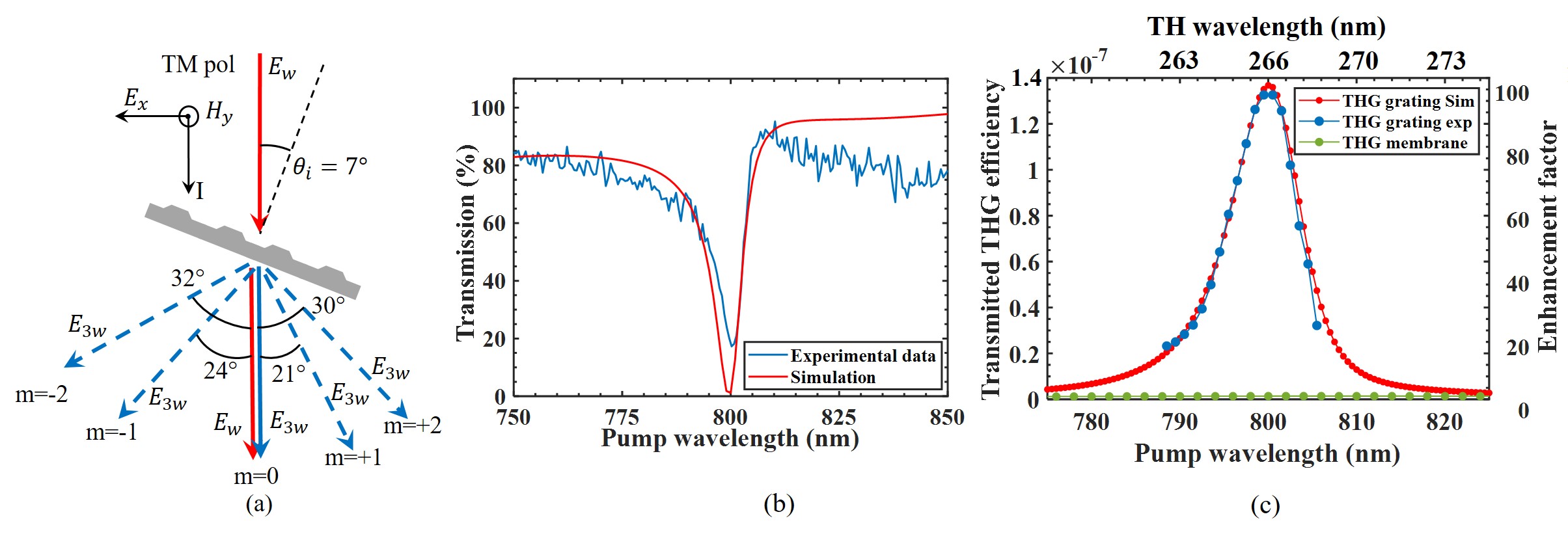}
    \caption{(a) Schematic of pump illumination configuration and THG diffraction orders; (b) Theoretical and  experimental transmittance indicating the resonance at 800 nm;  (c)  Theoretical and experimental THG efficiency (left axis) and enhancement (right axis) of the metasurface with respect to the unpatterned membrane.}
    \label{fig5}
\end{figure*}

The second set of nonlinear measurements was performed to investigate the resonant third-harmonic generation (THG) from the shallow-etched grating. After fabrication, the targeted guided-mode resonance of the shallow structure was found to be red-shifted to approximately 860 nm under normal-incidence TM illumination. By tilting the sample to an angle of incidence of 7° (see Supplementary Information, Fig. S3), a sharp resonance was observed at 800 nm [Fig.\ref{fig5}(b), blue curve]. The sample was therefore fixed at this angle for the subsequent nonlinear measurements. The corresponding numerical result is shown by the red curve. The observed feature corresponds to a guided-mode resonance of the grating, which enables strong localization of the electromagnetic field within the high-index bars. The resulting field confinement [Fig.\ref{fig1}(f)] leads to a resonant enhancement of THG, in agreement with both theoretical predictions and experimental observations. No resonant features are observed under TE-polarized illumination.

Fig.\ref{fig5}(c) presents the spectral dependence of the THG signal in the vicinity of the resonance. The left y-axis shows the absolute conversion efficiency from the pump to the third harmonic, while the right y-axis indicates the enhancement factor relative to the THG generated from a planar $Si_3N_4$ membrane measured under identical experimental conditions. The red curve represents the theoretical results obtained using COMSOL Multiphysics, a commercial finite-element method (FEM) solver. The simulations incorporate the previously retrieved wavelength-dependent datasets of the linear refractive index and nonlinear susceptibility. A peak conversion efficiency of $\eta_{TH}=\frac{P_{TH}}{P_{pump}}=1.4\ast 10^{-7}$ is predicted, corresponding to an enhancement factor of $\sim100$ compared to the THG from the unpatterned membrane (blue curve). This peak enhancement may also be interpreted as increase in the effective $\chi^{(3)}$ of the structure by a factor of 10, compared to the intrinsic $\chi^{(3)}$ of silicon nitride.

Given the grating periodicity of approximately 490 nm, the pump beam does not generate additional diffraction orders, while the designated TH around 266 nm will be emitted in several diffraction orders. Fig.\ref{fig5}(a) shows the zeroth and first two diffraction orders of the generated THG, which were experimentally detected as follows: the angle of incidence of the pump beam was fixed at 7° while the detection arm was moved horizontally in the plane of incidence, performing a full angular scan (+/- 90°). In this way we retrieved the first two diffraction orders of the generated TH, on each side of the transmitted zero order, as schematically represented in Fig.\ref{fig5}(a). The relative intensity of the different diffraction orders were recorded, and their sum helps us retrieve the total conversion efficiency to correctly corroborate our theoretical predictions, as in the COMSOL platform we simulate the total transmitted light passing through and integrated over the area of transmission. The corresponding experimental measurement of total THG as a sum of all diffracted orders, is represented in blue curve, showing a peak conversion efficiency of $\eta_{TH}=\frac{P_{TH}}{P_{pump}}=1.35\ast 10^{-7}$, corresponding to an enhancement factor of $\sim97$, or an increase in the effective $\chi^{(3)}$ of the structure by approximately a factor of 9.85 compared to the intrinsic $\chi^{(3)}$ of silicon nitride.

\begin{figure*}[hbt!]
    \centering
    \includegraphics[width=.9\linewidth]{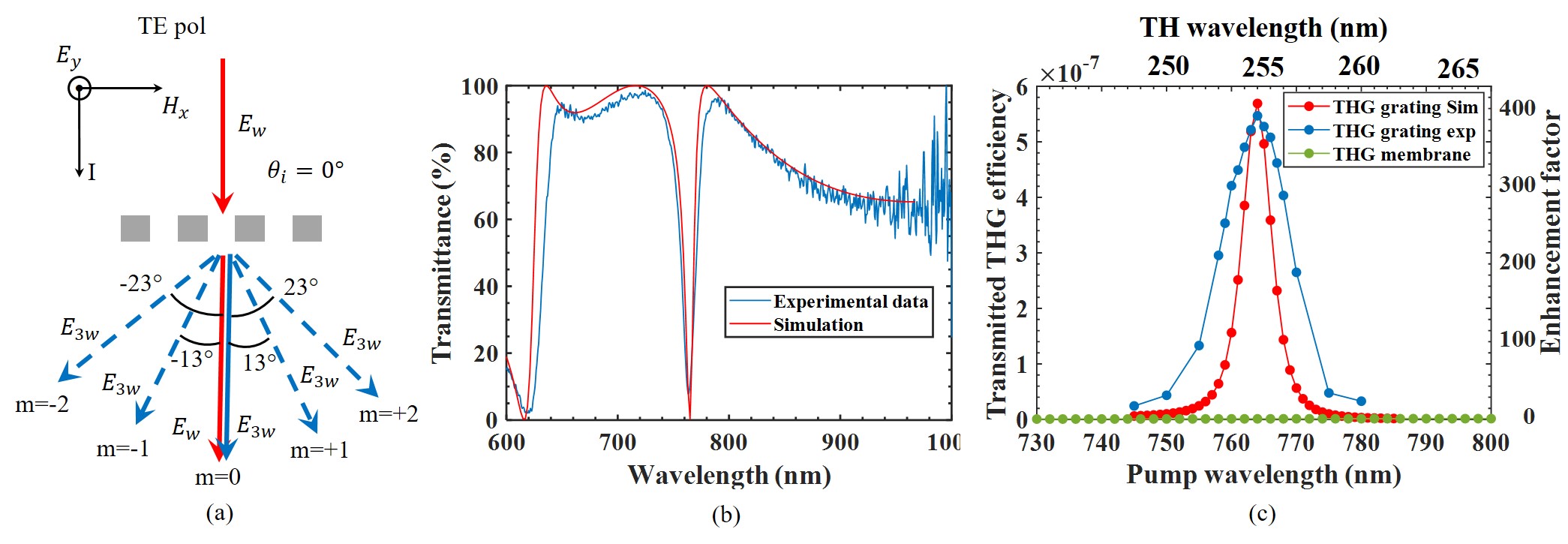}
    \caption{(a) Schematic of pump illumination configuration and THG diffraction orders; (b) Theoretical and experimental transmittance indicating the resonance near 764 nm;(c) Predicted and observed transmitted THG efficiencies (left axis) and enhancement (right axis) of the metasurface with respect to the unpatterned membrane.}
    \label{fig6}
\end{figure*}

The third set of nonlinear measurements are carried out to retrieve the resonant THG from the fully etched grating. After fabrication, this grating shows the characteristic Mie type resonance near 764 nm (Fig.\ref{fig6}(b)). The geometrical parameters h and d in the simulated structure are slightly modified to fit the experimental curve shown in blue, and the simulated curve is shown in red. Under TE illumination, the dip in the reflectance spectra comes as a result of strong light localization in the bulk of the material [Fig.\ref{fig1}(i)], and the corresponding resonant THG spectral variation, as theoretically predicted, is shown in Fig.\ref{fig6}(c). Like the previous structure, for a grating periodicity of approximately 500 nm, the pump beam does not generate additional diffraction orders, while its TH around 254 nm will produce several diffraction orders. Fig.\ref{fig6}(a) shows the zeroth and first two diffraction orders of the generated THG, which were experimentally detected in the following process: the angle of incidence of the pump beam was fixed at 0° while the detection arm was moved horizontally in the plane of incidence, performing a full angular scan (+/- 90) and we retrieved the first two diffraction orders of the generated TH, on each side of the transmitted zero order, as schematically represented in Fig.\ref{fig6}(a). The collection of the diffracted harmonics is once again necessary to correctly account for the total efficiency, as predicted by our theoretical predictions, both presented in Fig.\ref{fig6}(c). The relative intensity of the different diffraction orders were recorded and the sum of them gave us the total conversion efficiency of $\eta_{TH}=\frac{P_{TH}}{P_{pump}}=5.5\ast 10^{-7}$ with a peak enhancement factor of $\sim400$ times, with respect to the flat membrane, and is presented as the blue curve. This enhancement is equivalent to an increase of the effective $\chi^{(3)}$ of the structure by a factor of 20, compared to the intrinsic  $\chi^{(3)}$ of silicon nitride. The red curve, which is the theoretical prediction, is in excellent agreement with the experimental data. The slightly wider spectral response of the experimental curve may be attributed to slight departures from design parameters, like rounded nanowire corners instead of having perfectly square cross-section, slight differences in nanowire spacing and depth within the laser spot, as assumed in our simulations. These differences are apparent in both Figs.\ref{fig6}(b) and \ref{fig6}(c): both linear and THG bandwidths are wider compared to the simulations.

\section{Conclusion}
We have reported TH emissions from two different $Si_3N_4$metasurfaces, comparing their resonant properties in terms of field localization and generated signal. We have also compared their resonant THG efficiency to the efficiency of a flat $Si_3N_4$ membrane. Thanks to the extremely low losses of $Si_3N_4$ at the pump wavelength, TH conversion efﬁciency from the flat membrane is already higher than metal-based nanostructures, which is then further enhanced through resonant, polarization selective tunable metasurfaces. Using our theoretical model and simulations, we find excellent agreement with our experimental observations of both linear and third harmonic generation from the gratings, at visible and UV wavelengths. This comparison shows strong enhancement in the pump to THG conversion efficiencies well in excess of two orders of magnitude. Theoretical predictions show that simple geometrical rearrangements can significantly improve the THG conversion efficiencies to nearly three orders of magnitude, leaving open the possibilities of what we can achieve with advances in precision nano-fabrication techniques in the future. 

This study maps an important step towards understanding and showcasing $Si_3N_4$ metasurfaces as a promising candidate for highly efficient photonic nanostructures operable from the UV to the infrared regions. The successful generation and detection of higher-order harmonics depend on a detailed understanding of the high refractive index dielectric structures, its resonant properties, mode confinements and the NL electron dynamics at the nanoscale. Such insights are fundamental for developing coherent and efficient nanoscale UV/DUV/EUV light sources that facilitate miniaturization and integration into portable devices. Moreover, silicon nitride on a hybrid platform with III–V materials such as InP, opens up a whole new generation of applications and system-on-chip applications. Low optical losses, transparency from visible through the MIR, compatibility with CMOS and wafer-scale foundry processes, and high-power handling capabilities are among the key attributes of these linear and nonlinear PICs. A wide array of applications will beneﬁt from the optical transparency and low loss from the visible out to the IR, including optical inertial rotation sensors, microwave synthesizers, quantum communications, biophotonics, and nanoparticle analysis applications. The high spectral tunability of the structures also establish $Si_3N_4$ metasurfaces as a perfect candidate for system-on-chip PICs that operate over an unprecedented optical bandwidth range, from the visible wavelength  ($\sim400 nm$) out to beyond the infrared ($>2.3\mu m$) and deliver performance previously only achievable with bulk optic technologies.

\begin{acknowledgments}
SM, JT, and CC acknowledge the Spanish Agencia Estatal de Investigación from Ministerio de Ciencia, Investigación y Universidades (project PID2023-148620NB-I00 granted by MICIU/AEI/10.13039/501100011033 and FEDER, UE). S. M. acknowledges predoctoral grant FPI-UPC 2022, funded by Universitat Politècnica de Catalunya (Government of Catalunya). 
\end{acknowledgments}

\section*{Data Availability Statement}

The data that support the findings of this study are available from the corresponding author upon reasonable request.
\appendix

\section{Appendixes}
\subsection{Microscopic hydrodynamic model for nonlinear optics at nanoscale}
The nonlinear light-matter interactions in the thin film membrane are simulated using a theoretical model described in detail in references [39-42]. In the case of $Si_3N_4$ we do not consider contributions from conduction electrons or holes because pumping occurs at wavelengths that correspond to energies below what is required to excite valence electrons into the conduction band. The equations of motion and the process of retrieval of the material properties of $Si_3N_4$ are described here:
 
\begin{widetext}
 \begin{equation}
{\mathbf{\Ddot{P}}}_{bj} + \Tilde{\gamma}_{bj} {\mathbf{\Dot{P}}}_{bj} + \Tilde{\omega}_{0,bj}^2 \mathbf{P}_{bj} - \Tilde{\beta}_j (\mathbf{P}_{bj} \mathbf{\cdot} \mathbf{P}_{bj}) \mathbf{P}_{bj}+ \Tilde{\Theta}_j (\mathbf{P}_{bj} \mathbf{\cdot} \mathbf{P}_{bj})^2 \mathbf{P}_{bj} - \Tilde{\Delta}_j (\mathbf{P}_{bj} \mathbf{\cdot} \mathbf{P}_{bj})^3 \mathbf{P}_{bj}= \frac{n_{0,b} e^2 \lambda_0^2}{m_{bj}^* c^2} \mathbf{E} + \frac{e \lambda_0}{m_{bj}^* c^2} (\mathbf{P}_{bj} \mathbf{\cdot} \nabla)\mathbf{E} + \frac{e \lambda_0}{m_{bj}^* c^2} {\mathbf{\Dot{P}}_{bj} \times \mathbf{H}}
\label{eqn2}
\end{equation}
   
\end{widetext} 

The equation above is a scaled equation of motion for bound electrons. The subscript $b$ stands for bound, and the counter $j$ represents the $jth$ bound oscillator species.

Eq.(\ref{eqn2}) represents an expanded hydrodynamic model that includes a description of surface ($\mathbf{P}_{bj} \mathbf{\cdot} \nabla)\mathbf{E}$), magnetic ($\mathbf{\Dot{P}}_{bj} \times \mathbf{H}$) and bulk ($- \Tilde{\beta}_j (\mathbf{P}_{bj} \mathbf{\cdot} \mathbf{P}_{bj}) \mathbf{P}_{bj}+ \Tilde{\Theta}_j (\mathbf{P}_{bj} \mathbf{\cdot} \mathbf{P}_{bj})^2 \mathbf{P}_{bj} - \Tilde{\Delta}_j (\mathbf{P}_{bj} \mathbf{\cdot} \mathbf{P}_{bj})^3 \mathbf{P}_{bj}$) nonlinearities triggered by bound electrons. 

The constants $\Tilde{\beta}_j,\Tilde{\Theta}_j$, and $\Tilde{\Delta}_j$  are real, known, and depend on linear oscillator parameters like resonance frequency and lattice constant. $n_{0,b}$ is the bound electron density, in the absence of an applied field; $e$ is the electronic charge; $\lambda_0=1\mu m$ is a conveniently chosen scaling wavelength; c is the speed of light; $m^*_{bj}=m_e$  is the bound electron effective mass, for simplicity assumed to be equal to the free electron mass; $\xi = \frac{z}{\lambda_0}, \varsigma = \frac{y}{\lambda_0}, \zeta = \frac{x}{\lambda_0}, \tau = \frac{ct}{\lambda_0}$ are the scaled space and time coordinates that we used to calculate spatial and temporal derivatives. Finally, plasma frequencies and damping coefficients are scaled as follows: $\Tilde{\omega}^2_p = \frac{4 \pi n e^2}{m^*}\frac{\lambda^2_0}{c^2}, \Tilde{\gamma}=\gamma\frac{\lambda_0}{c}$. For additional, specific details about the model and method of solution, we refer the reader to references \cite{bib42,bib43,bib44,bib45}. 

The above mentioned method allows us to identify, distinguish, and explain the different nonlinear contributions to the harmonics generated by silicon nitride at the nanoscale at visible and UV wavelengths, to fit the THG characteristics of the planar membrane and to retrieve the wavelength dependent datasets of $\chi_\omega^{(3)}$ and $\chi_{3\omega}^{(3)}$, for our pump intensity and fixed laboratory environment.  

\subsection{COMSOL FEM Simulations}
The wavelength dependent local dielectric response (from Eq.\ref{eqn1}) and the wavelength dependent datasets of $\chi_\omega^{(3)}$ and $\chi_{3\omega}^{(3)}$ were introduced as $Si_3N_4$ material properties into the COMSOL platform, to conduct nonlinear simulations of the linear transmission and the nonlinear THG from the structured materials. The linear transmission and E and H field localization maps are solved by the Maxwell’s equations for E and H field propagation. For THG solutions, the nonlinear polarization term is introduced into the Maxwell's equations as follows:
  \begin{equation}
      \nabla\times\mu^{-1}_r(\nabla\times E)-k^2_0(\epsilon_r-\frac{j \sigma}{\omega\epsilon_0})E=\omega^2\mu_0\sum_i P_i
      \label{eqn3}
  \end{equation}
  
The following two third order nonlinearities were introduced as the nonlinear polarization term into the above mentioned EM wave equation.

1.	The Kerr effect in the wave equation of the pump wavelength is defined as follows:
\begin{equation}
    P_\omega = \chi^{(3)}_\omega E_\omega + 3 \epsilon_0 \chi^{(3)}_\omega  |E_\omega|^2 E_\omega
    \label{eqn4}
\end{equation}

2.	Bulk third order susceptibility in the THG wavelength is defined as follows:
  \begin{equation}
    P_{3\omega} = \chi^{(3)}_{3\omega} E_\omega + \chi^{(3)}_{3\omega}  (E_\omega \cdot E_\omega) E_\omega
    \label{eqn5}
\end{equation}

\nocite{*}
\bibliography{aipsamp}

\end{document}